\documentclass{emulateapj}

\newcommand{\gsim}{\mbox{$_>\atop^{\sim}$}}

\newcommand{\mum}{$\,\mu$m}

\newcommand{\arcs}{$^{\prime\prime}$}

\newcommand{\spitzer}{\textsl{Spitzer}}
\newcommand{\iras}{\textsl{IRAS}}
\newcommand{\iso}{\textsl{ISO}}

\newcommand{\boo}{Bo\"otes}
\newcommand{\ha}{H$\alpha$}
\newcommand{\kps}{\,km\,s$^{-1}$}

\newcommand{\boofn}{MIPS\,J142824.0+352619}
\slugcomment{Submitted to the Astrophysical Journal}

\shorttitle{A Hyperluminous Starburst Galaxy at $z=1.325$}
\shortauthors{Borys et al.}

\begin{document}

\title{\boofn: A Hyperluminous Starburst Galaxy at $z=1.325$}

\author{
C.\ Borys,$\!$\altaffilmark{1}
A.~W.\ Blain,$\!$\altaffilmark{1}
A. Dey,$\!$\altaffilmark{2}
E. Le Floc'h,$\!$\altaffilmark{3,4}
B.~T. Jannuzi,$\!$\altaffilmark{2}
V. Barnard,$\!$\altaffilmark{5}
C. Bian,$\!$\altaffilmark{1}
M. Brodwin,$\!$\altaffilmark{6}
K. Men\'endez-Delmestre,$\!$\altaffilmark{1}
D. Thompson,$\!$\altaffilmark{1}
K. Brand,$\!$\altaffilmark{2}
M.~J.~I. Brown,$\!$\altaffilmark{7}
C.D.\ Dowell,$\!$\altaffilmark{1,6}
P. Eisenhardt,$\!$\altaffilmark{6} 
D. Farrah,$\!$\altaffilmark{8,9}
D.T.\ Frayer,$\!$\altaffilmark{10}
J. Higdon,$\!$\altaffilmark{8}
S. Higdon,$\!$\altaffilmark{8}
T. Phillips,$\!$\altaffilmark{1} 
B.~T. Soifer,$\!$\altaffilmark{1,10}
D. Stern,$\!$\altaffilmark{5}
D. Weedman$\!$\altaffilmark{8}
}
\email{borys@caltech.edu}
\altaffiltext{1}{Caltech, 1200 E California Blvd., Pasadena, CA 91125}
\altaffiltext{2}{National Optical Astronomy Observatory, 950 North Cherry Avenue, Tucson, AZ 85719, USA}
\altaffiltext{3}{Steward Observatory, University of Arizona, 933 North Cherry Avenue, Tucson, AZ 85721}
\altaffiltext{4}{Observatoire de Paris, GEPI, 92195 Meudon, France}
\altaffiltext{5}{Joint Astronomy Center, 660 N, A'ohoku Place, Hilo, HI 96720, USA}
\altaffiltext{6}{Jet Propulsion Laboratory, 4800 Oak Grove Drive, Pasadena, CA, 91109}
\altaffiltext{7}{Princeton University Observatory, Peyton Hall, Princeton, NJ 08544}
\altaffiltext{8}{Astronomy Department, Cornell University, Ithaca, NY 14853, USA}
\altaffiltext{9}{Infrared Processing and Analysis Center, California Institute of Technology, Pasadena, CA, 91125}
\altaffiltext{10}{Spitzer Science Center, California Institute of Technology, Pasadena, CA 91125}

\begin{abstract}
Using the SHARC-II camera at the Caltech Submillimeter Observatory to obtain 350\mum\ images of sources detected with the MIPS instrument on \spitzer, we have discovered a remarkable object at $z=1.325\pm0.002$  with an apparent Far-Infrared luminosity of ${\rm 3.2 (\pm0.7)\times10^{13}\,L_\odot}$.   Unlike other $z>1$ sources of comparable luminosity selected from mid-IR surveys, \boofn\ lacks any trace of AGN activity, and is likely a luminous analog of galaxies selected locally by \iras, or at high redshift in the submillimeter.  This source appears to be lensed by a foreground elliptical galaxy at $z=1.034$, although the amplification is likely modest ($\la10$).  We argue that the contribution to the observed optical/Near-IR emission from the foreground galaxy is small, and hence are able to present the rest-frame UV through radio Spectral Energy Distribution of this galaxy.  Due to its unusually high luminosity, \boofn\ presents a unique chance to study a high redshift dusty starburst galaxy in great detail.
\end{abstract}


\section{Introduction}
The \iras\ satellite discovered a population of dusty Ultraluminous Infrared Galaxies \citep[ULIRGS;L$_{\rm IR}>10^{12}$L$_\odot$;][]{1987ApJ...320..238S,1988ApJ...325...74S}  in the local Universe that we now know to be significant at high redshift as well. Followup studies of sub-millimeter galaxies (SMGs) found in SCUBA and MAMBO surveys \citep[e.g.][]{1997ApJ...490L...5S,1999ApJ...515..518E,2002MNRAS.331..817S,2003MNRAS.344..385B,2004MNRAS.354..779G} have shown that ULIRG activity peaks near $z\sim2.2$ \citep{2005ApJ...622..772C}, with a space density about $10^3$ times higher than locally.  While these distant galaxies play a central role in the course of galaxy evolution, limited photometry at wavelengths where they emit most strongly restricts detailed characterization of their stellar, nuclear, dust, and evolutionary properties.

This is due in large part to a lack of sensitivity by instruments operating at Mid/Far-IR wavelengths. While \iras\ and \iso\ were able to detect a number of \mbox{$z\ga1$} sources, many fall into the {\it Hyper}-luminous regime (HLIRGS; L$_{\rm IR}>10^{13}$L$_\odot$), of which a notable example is IRAS-F15307+3252 \citep[z=0.93;][]{1994ApJ...424L..65C}. This high luminosity tail is dominated by systems with an active galactic nucleus (AGN) \citep[e.g.][]{1999ApJ...522..113V,2001ApJ...552..527T}, but there may be a selection effect at work here; the fraction of hot dust due to an AGN can be much higher than in starbursting galaxies, meaning that the emission peaks at shorter wavelengths, closer to the \iras\ bands.  Indeed, followup studies of these systems with SCUBA have shown them to be faint or invisible at long sub-mm wavelengths \citep{2001MNRAS.326.1467D,2002MNRAS.335.1163F}. In contrast, sources at higher redshift selected in the sub-mm by SCUBA, although harboring an AGN, probably have extreme luminosities driven by intense star-formation \citep{alexanderb}.  Again this may be due to a selection effect, since hot AGN at high redshift will be biased away from the long-wavelength SCUBA channels.

In both cases, the luminosities and dust temperatures are based on assumptions on the shape of the Spectral Energy Distribution (SED), which are drawn from a handful of bright \iras\ selected galaxies in the nearby Universe (and hence not necessarily representative of high$-z$ galaxies).  Altogether, we are left with an incomplete picture of luminous dusty sources between $z\sim1-2$, where \iras\ and \iso\ detected sources taper off, and before SCUBA sources peak.  The problem is compounded by the fact that many such sources lie in the so-called `redshift desert', which optical spectrometers have difficulty probing.

One goal of the {\it Spitzer Space Telescope} \citep{2004ApJS..154....1W} is to fill in this redshift interval with unbiased samples of luminous IR galaxies.  Combining the increased sensitivity afforded by the Multi-band Imager for \spitzer\ \citep[MIPS;][]{2004ApJS..154...25R} with wide area Legacy and GTO imaging surveys, we now have large samples of Far-IR selected galaxies.  Ground-based sub-mm observations in the 350/450\mum\ windows are a particularly valuable complement, since at \mbox{$z\gsim1$}, they probe at or near the peak of the SED (typically at rest frame $\sim100$\mum), allowing tighter constraints on luminosity and dust temperature. To that end, we have initiated a program of 350\mum\ sub-mm follow-up of MIPS selected sources using the SHARC-II camera \citep{sharc} at the Caltech Submillimeter Observatory (CSO).  Here we report a striking detection of a starburst dominated HLIRG at $z=1.325$.

In the discussion that follows, any calculation requiring cosmology assumes $\Omega_{\rm M}=0.27, \Omega_\Lambda=0.73$, and H$_0$=71\,km\,s$^{-1}$\,Mpc$^{-1}$.

\section{Field Description and Selection Criteria}
Through a \spitzer\ GTO program (PI: Soifer), 24, 70 and 160\mum\ images of the NOAO Deep Wide-Field Survey \citep[NDWFS;][]{1999ASPC..191..111J} \boo\ field were obtained using the MIPS camera in February 2004.   The 9.3\,deg$^2$ survey reached $1\sigma$ sensitivity levels of roughly 0.1, 7, and 15\,mJy at 24, 70, and 160\mum\ respectively.

This field boasts an impressive collection of multi-wavelength data:  radio imaging for part of the field, including the region around the source presented here, is available from a moderately deep ($1\sigma\sim15\mu$Jy) and high angular resolution (1.5\arcs\ FWHM) VLA 1.4\,GHz observation \citep{2005ApJ...626...58H}. Another GTO program surveyed the entire field with the IRAC camera, providing deep imaging at 3.6, 4.5, 5.8, and 8.0\mum\ \citep{2004ApJS..154...48E}.  Combined with the NDWFS optical data, and our own $JHK$ photometry obtained from the WIRC camera at Palomar for the target discussed here, the wavelength coverage is sufficient to fully characterize the rest frame $B-$ through $K-$ band SED up to $z\sim3$.  The third \spitzer\ GTO component to the \boo\ observations involves IRS spectroscopy of 24\mum\ selected sources (PI: Weedman), which included the object presented here.

To select sources for 350\mum\ SHARC-II observations, we used a list of MIPS sources that were detected in all three bands and were confirmed to have unambiguous optical counterparts. We further culled the list by selecting only those that appeared compact and red in the optical images (compared to the field). These additional conditions are expected for SMGs at higher redshift \citep[e.g.][]{2004MNRAS.355..485B}.   For  commonly used ULIRG SEDs (i.e. Arp 220 and HR10), $S(350\mu{\rm m})/S(160\mu{\rm m})$ varies from $\sim0.3$ at $z=1$ to $\sim1.0$ at $z=2$.  Our goal was to select $z=1-2$ targets that could be detected at $>4\sigma$ confidence with SHARC-II in under 1 hour of good weather conditions, and hence we used a flux cut of $S(160\mu{\rm m})>200$\,mJy.  These conditions imply that any source found should have a luminosity near $10^{13}$L$_\odot$.   This paper concentrates on the target from this list which has the brightest 160\mum\ flux: \boofn.  Although the NDWFS standard is to use the $R$ band to derive the position and name\footnote{The NDWFS catalogs, and a description of the naming conventions are available at \texttt{http://www.noao.edu/noao/noaodeep/index.html}}, we use the $I-$band image which has the highest signal-to-noise ratio of all the available data on this object.  For completeness, the J2000 coordinates are: 14$^{\rm h}$28$^{\rm m}$24\fs07 +35\degr26\arcmin19\farcs4.

\section{Targeted observations of \boofn}
\subsection{Sub-mm imaging and photometry}
\begin{figure}
\plotone{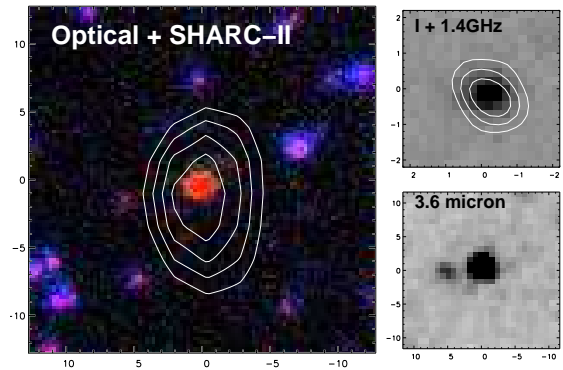}
\caption{The main panel shows SHARC-II 350\mum\ flux contours at [3.0,3.5,4.0,4.5]$\sigma$ overlaid on a 30\arcs$\times$30\arcs\ square NDWFS false color image of the field.  Axes are aligned along cardinal directions, with North facing up. The lower of the two smaller cutouts shows the IRAC 3.6\mum\ image of a region around \boofn.  The upper panel presents a closeup view of the $I-$band image with VLA contours at [0.2,0.3,0.4] mJy levels.  Although there at least three faint red sources in the vicinity of \boofn, the radio position clearly favors the central bright object.}
\label{fig:booimg}
\end{figure}

Observations using the SHARC-II camera were carried out at the CSO on UT 2004 June 06. Atmospheric opacity was low, with $\tau_{\rm 350{\mu}m} = 1.7$. Pointing and flux calibration were performed on the nearby source Arp~220, and data reduced using the {\sc crush} software package \citep{crush}. Instrumental fluxes were measured within 20\arcs\ apertures (corresponding to the $3\sigma$ width of the SHARC-II beam), and scaled by a factor derived from the same aperture on the calibration images.  In Fig.~\ref{fig:booimg}, we plot the 350\mum\ contours on a false-color optical image.

The SHARC-II detection probes near the peak of the SED, but to constrain its shape further into the Rayleigh-Jeans portion of the spectrum, longer wavelength observations are required. On UT 2004 August 28, one hour of photometry data at both 850 and 450\mum\ were obtained in good weather ($\tau_{850\mu{\rm m}}=0.3$) using SCUBA \citep{scuba} on the James Clerk Maxwell Telescope (JCMT).  The data were reduced using {\sc surf} \citep{surf} and flux calibrated against CRL 2688.  Since the source was observed while setting late in the semester, these data were obtained before and during sunset when calibration uncertainties at 450\mum\ (due to the sensitivity of the dish to temperature) render that channel's data unusable.  At 850\mum, this is not a problem, and the measured flux is accurate to about 5\%.

\subsection{Spectroscopic observations}
\begin{figure}
\plotone{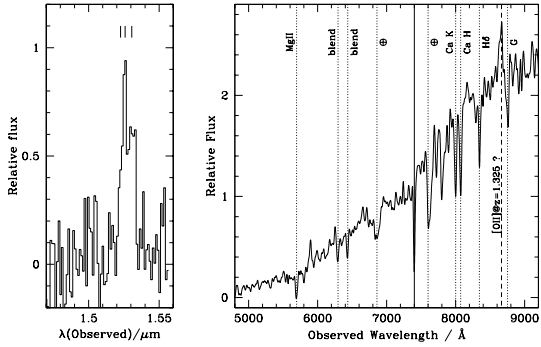}
\caption{LEFT: NIR emission lines detected at Keck using the NIRSPEC instrument on UT 2005 March 17. The exposure time was 2400s, and used the moderate resolution setting with a 0\farcs76 slit. To improve the signal-to-noise ratio, the spectrum is smoothed to 0.02\mum\ bins. Vertical lines denote the positions expected for \ha\ and the flanking [NII] lines at $z=1.325$. RIGHT: Keck-DEIMOS spectra of \boofn, smoothed to highlight several absorption features which suggest the presence of a $z=1.034\pm0.002$ foreground galaxy.  The thick line at 7400\AA\ denotes the divide between the red and blue arms on the spectrometer.  The A and B band telluric lines are denoted by the symbol $\oplus$.  We also indicate where [OII] would appear at $z=1.325$.  Although this does appear to coincide with an emission line nearby, the difficulty in removing atmospheric lines at the red end renders it inconclusive. }
\label{fig:spec}
\end{figure}

\citet{vandana} present the  \spitzer\ IRS spectrum for \boofn, and determine a redshift of $z=1.34\pm0.02$.  Despite the many strongly detected PAH features, the relatively low resolution of the spectrometer prevents the determination of a more accurate redshift.  With this spectrum as a guide, we used the NIRSPEC instrument on Keck II to search for \ha.   Although the conditions were not photometric (overcast skies and a seeing of $\sim1$\arcs), we are able to clearly detect it (Fig.~\ref{fig:spec}) and derive a redshift of $z=1.325\pm0.002$ by fitting Gaussians jointly to \ha\ and the flanking [NII] lines.   The \ha\ line width is $530\pm160$\kps\ (FWHM), but we caution that this may be an overestimate since it may contain contributions from the [NII] lines which were difficult to de-convolve due to the low signal-to-noise of the spectrum.

While the Near- and Mid-IR spectroscopy find an object at $z=1.325$, optical observations uncover something else.  A spectrum taken with the Keck-DEIMOS instrument on UT 2005 May 06 reveals the presence of a $z=1.034\pm0.002$ galaxy identified by MgII(2800), the Calcium H and K lines, as well as common blended features (see Fig.~\ref{fig:spec}).  These features suggest it is an elliptical galaxy.  No line features from the $z=1.325$ galaxy are seen, although there is generally positive flux near $8669$\AA\, where [OII](3727) would be expected.  However, the spectrum is much noisier on the redder end, and we cannot say with confidence whether this feature is a real line, or simply a noise excursion.  The $z=1.034$ galaxy is likely lensing the background source, and is a complication we address in \S\ref{sec:lensing}.

\section{Determining the SED of \boofn}
\begin{figure}
\plotone{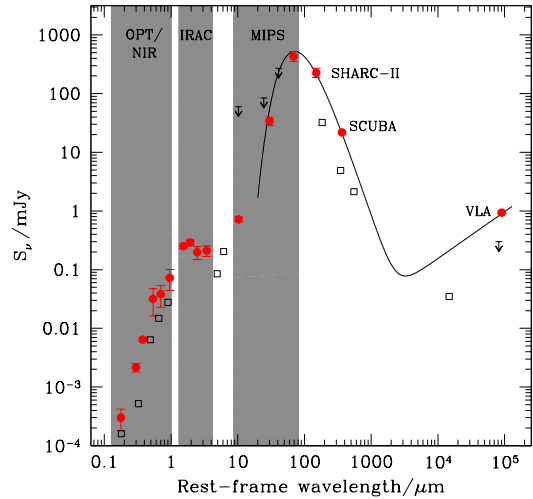}
\caption{The SED of \boofn, shifted to the rest frame at $z=1.325$.  Overlaid on the MIPS+sub-mm points is the best fit grey-body.  The radio portion of the SED assumes a spectral index of 0.7, which is commonly used to describe starburst galaxies \citep{1992ARA&A..30..575C}.  Squares and upper limits represent the rest-frame SED of HR10.}
\label{fig:sed}
\end{figure}

We present a summary of the measured photometry in Table~\ref{tab:photom} and Fig.~\ref{fig:sed}.  Concentrating on the Far-IR portion, we find that the MIPS, SHARC-II, and SCUBA observations are well fit by a modified blackbody of the form \mbox{$B(\nu,T/(1+z))\nu^\beta$}, where \mbox{$B(\nu,T/(1+z))$} is the usual Planck function, and $z,T,\beta$ describe the redshift, dust temperature and emissivity respectively.  The $T/(1+z)$ degeneracy \citep[e.g.][]{2003MNRAS.338..733B} reflects the fact that a hot dust source at high redshift has the same SED shape as a colder one at lower redshift. From our multi-wavelength spectroscopy, it is clear that the Far-IR emission is coming from the $z=1.325$ galaxy, and hence we are able to derive a dust temperature of $T=42.7\pm2.7$ and emissivity $\beta=1.5\pm0.2$.  The Far-IR luminosity, integrated between rest-frame 8 and 1000\mum\ is ${\rm L = 3.2 (\pm0.7)\times10^{13}\,L_\odot}$, placing it squarely in the hyperluminous classification.  Of course, this could be complicated by lensing, which we now discuss.

\begin{deluxetable}{lcl}
\tabletypesize{\scriptsize}
\tablecaption{Multiwavelength photometry of \boofn.}
\tablehead{
\colhead{Wavelength} & \colhead{Flux} & 
\colhead{Instrument\tablenotemark{a}}
}
\startdata
445\,nm$(B_W)$ & $0.25\pm0.03\,\mu$Jy & MOSAIC-1  \\
658\,nm$(R)$   & $2.02\pm0.08\,\mu$Jy & MOSAIC-1  \\
806\,nm$(I)$   & $6.21\pm0.11\,\mu$Jy & MOSAIC-1  \\
1.22\mum$(J)$  & $31.8\pm5.2\,\mu$Jy  & WIRC  \\
1.63\mum$(H)$  & $38.4\pm5.2\,\mu$Jy  & WIRC  \\
2.19\mum$(K)$  & $72.4\pm9.4\,\mu$Jy  & WIRC  \\
3.6\mum\  & $250.9\pm7.5\,\mu$Jy    & IRAC \\
4.5\mum\  & $290.6\pm8.7\,\mu$Jy    & IRAC \\
5.8\mum\  & $198.3\pm16.6\,\mu$Jy   & IRAC \\
8.0\mum\  & $211.2\pm14.2\,\mu$Jy   & IRAC \\
24\mum\   & $0.72\pm0.07$\,mJy      & MIPS    \\
70\mum\   & $34\pm6$\,mJy           & MIPS    \\
160\mum\  & $430\pm90$\,mJy         & MIPS    \\
350\mum\  & $226\pm45$\,mJy         & SHARC-II\\
850\mum\  & $21.9\pm1.3$\,mJy       & SCUBA   \\
20cm      & $0.937\pm0.039$\,mJy    & VLA     \\
\enddata
\tablecomments{Optical fluxes are taken from directly the NDWFS catalog, and are measured in 5\arcs\ apertures.  IRAC fluxes are measured using the same sized aperture, and and are corrected to total magnitudes by using the IRAC PSF.}
\tablenotetext{a}{MOSAIC-1 is the wide-field optical imager on the Mayall 4m at  KPNO, and WIRC is the NIR camera on the Palomar 5m.}
\label{tab:photom}
\end{deluxetable}

\subsection{Lensing of \boofn}\label{sec:lensing}
The radio emission should be a good proxy to the location of the Far-IR emission \citep[e.g.][]{2000AJ....119.2092B,2001ApJ...548L.147C,2002MNRAS.337....1I}, and we find that the peak of the radio emission is only 0\farcs1$\pm$0\farcs3 away from the $I-$band centroid. Furthermore, \citet{iono} use the SMA to obtain a $\sim2$\arcs\ resolution image of this source at 890\mum, and find that it is within 0\farcs4$\pm$0\farcs3 of the optical position. Hence we argue that the alignment between the source and lens is very strong. Adding the VLA/SMA uncertainty in quadrature with the optical astrometric accuracy, we estimate that the observed positional offset between the source and lens is $\la$0\farcs5. 

The foreground elliptical is not the source of the Far-IR emission, but how much does it contribute in the observed optical--Near-IR?  \boofn\ is an extremely red object (ERO), having an $R-K$ color of $5.9\pm0.2$ (Vega).  Since the ERO criterion matches the observed colors of ellipticals at $z\sim1$, and since the DEIMOS spectrum is consistent with a K+A galaxy, it is possible that the foreground $z=1.034$ elliptical dominates the observed Near-IR emission.  However, bright submm galaxies can exhibit similar colors due to extreme extinction \citep[e.g.][]{2002ApJ...577L..83W,2004AJ....127..728F,2004ApJ...605..645W}, and hence the lensed source is also a good candidate.  

To see if we can place constraints on the relative contributions of the source and lens, we employ a Singular Isothermal Sphere (SIS) to represent the foreground elliptical.  In this simple model, the relevant angular scale is roughly twice the Einstein radius ($2\theta_E$), which is the separation between the two images of the lensed source (in the case where the {\em unlensed} separation between the source and lens is $<\theta_E$) or between the lens and the single image of the source which results when the unlensed separation is $>\theta_E$.  The $\sim2$\arcs\ resolution of the radio and SMA images makes it impossible to distinguish between the two lensing scenarios, but we can use the 0\farcs5 upper limit between the SMA/radio position and the optically detected source as a limit on $2\theta_E$.  This in turn places a limit on the velocity dispersion of the lensing elliptical: the SIS model gives $\theta_E=0$\farcs$23(\sigma_L/220$\kps$)^2$ and hence we derive  $\sigma_L<230$\kps.

Using the Near-IR Faber-Jackson scaling relation (Fig.~11 of \citet{k20}), the absolute rest-frame $K-$band magnitude for an $z\sim1$ elliptical with $\sigma_L=230$\kps\ is M$_K=-26.0\pm0.2$.  In flux units, this corresponds to 33$\mu$Jy.  Since the rest-frame $K-$band at $z=1$ corresponds directly to the IRAC 4.5\mum\ channel, we see (via Table~\ref{tab:photom}) that the foreground elliptical contributes significantly less in the Near-IR than the background source, by at least a factor of $\sim6$.

This agrees, at least qualitatively, with the observed optical-IRAC SED, which does not show strong evidence of being composed of multiple objects.  Using the photometric redshift code developed by the \boo-IRAC team \citep{brodwin}, we obtain an estimate of $z_{\rm phot}=1.44^{+0.19}_{-0.15}$, in good agreement with the spectroscopic result for the background object.  Note however, that the absorption lines seen in the optical spectrum are rather deep, hence the foreground galaxy is likely dominating the emission shortward of $\sim1$\mum.  This would suggest that the background source is even redder than what is observed.

The faint red source 5\arcs\ east of \boofn\ (see Fig.~\ref{fig:booimg}) could not be a second image of the source under the SIS model since the brighter image (when two are present) is furthest from the lens.  However, we note that the three highest signal-to-noise images, $I$, 3.6\mum, and 4.8\mum, demonstrate a flux ratio of 9, 10, and 11 respectively between \boofn\ and this companion.  The uncertainty for each ratio is $\sim1.0$, meaning that it is conceivable that the objects have the same colors (and by inference are the different images of the same lensed source).  Deeper imaging and spectroscopy would be required to verify this.

\subsubsection{Estimating the lensing amplification}
Although the well constrained Far-IR SED provides an accurate {\em apparent} luminosity, the value of the lensing amplification, $\mu$, is needed to derive its true brightness.  For an ideal point source, $\mu$ varies from being infinite if the source is directly behind the lens, to $\mu\sim2$ at the 0\farcs5 upper limit to the observed angular separation between source and lens.
However, \boofn\ is not a point source.  We can use the Stefan-Boltzmann law to estimate its physical size assuming that the object is a spherical blackbody.  In practical units, the diameter of the galaxy using this relation is ${\rm D/kpc =  0.94(L/10^{12}L_\odot)^{0.5}({T/40\,K})^{-2}}$. This results in a source size of $4.6\mu^{-0.5}$\,kpc (0\farcs5$\mu^{-0.5}$), where we have now included the unknown lensing factor explicitly in the luminosity.  \citet{1993ApJ...414L..13D} used this approach to estimate source sizes on the order of 400\,pc for typical \iras\ selected ULIRGs.  If \boofn\ is inherently similar, then the lensing must be very strong, with $\mu\sim100$.  However, even 400\,pc corresponds to an non-negligible fraction of the Einstein radius (400pc = 47\,mas at $z=1.325$). This argues against a high lensing factor for \boofn. 

Furthermore, ULIRG sizes on kpc scales are not uncommon.  \citet{2004ApJ...611..732C} used high resolution radio imaging to show that $z\sim2.2$ SCUBA selected galaxies are extended over scales as large as 4\,kpc.  If \boofn\ is indeed extended, then the overall lensing amplification is low, since the bulk of the emission would be due to parts of the galaxy that are not significantly lensed.    In such a scenario, differential lensing would affect the observed properties of the source.  For instance, the stars and dust may dominate in distinct parts of the galaxy, meaning that the overall amplification for each component may be different.  However, the Far-IR and radio emission should arise from the same parts of the galaxy, so it is reasonable to assume a common lensing amplification and use those flux estimates to characterize the dust properties of the background source.

\section{Discussion}
The source of energy powering Far-IR luminous objects is typically attributed to AGN or starburst activity, and in this section we use the available data to determine the relative contributions from both processes to \boofn.

Our $\sim43$K dust temperature estimate is comparable to the mean $36$K derived via sub-mm observations of nearby \iras\ selected star-forming galaxies \citep{2000MNRAS.315..115D}; the dust temperatures attributed to AGN are much larger ($>60$K).   Furthermore, \citet{vandana} find that the mid-IR spectra is dominated by PAH features, which usually indicate a star-burst power source since they are easily destroyed by the hard radiation field of an AGN \citep[e.g.][]{1998ApJ...505L.103L}.  Finally, we find that the `q' parameter \citep{1992ARA&A..30..575C} which relates the Far-IR and radio luminosity agrees well with values derived for star-forming galaxies in the local Universe : \citet{1992ARA&A..30..575C} derives $2.3\pm0.2$ for local galaxies while \boofn\ has $q=2.5\pm0.2$.

If the observed optical/Near-IR flux is indeed dominated by the $z=1.325$ source, we note that the SED lacks any prominent break near 4000\AA.  This suggests that the optical spectrum is due to a young ($\la50\,$Myr) obscured stellar component, and hence supports the star-formation case.  It would also be further evidence against a strong AGN component, as the stellar bump is clearly seen near 1.6\mum\ (rest-frame).

Altogether it seems much more likely that the source is powered by vigorous star-formation.   These results do not preclude the presence of an AGN in \boofn, but only suggest that starburst activity dominates the Far-IR properties of the source.  Using our derived Far-IR luminosity and adopting the relation in \citet{1998ARA&A..36..189K}, we estimate a star formation rate of SFR$=5500\pm1000\,\mu^{-1}$M$_\odot$yr$^{-1}$.  Note that the empirical luminosity-temperature (L-T) relationship presented in \cite{2003MNRAS.338..733B} for sub-mm galaxies shows that the average luminosity for a source of 43K is $\sim10^{12}$L$_\odot$.  If \boofn\ obeys this trend, then the inferred lensing amplification is $\mu\sim10$, and the SFR still at least several hundred solar masses per year.

\section{Summary}
At $z=1.325$, \boofn\ allows us to gain insight on the population of luminous dusty galaxies between $z=1-2$.    The agreement with the radio/Far-IR correlation and the lack of broad or high excitation spectral lines suggests that \boofn\ is a starburst-dominated galaxy.  Although its observed extreme luminosity is likely enhanced from lensing by a foreground galaxy at $z=1.034$, the amplification is likely not higher than $\sim10$.  Hence \boofn\ is a bright ULIRG not dissimilar to those selected by \iras\ in the local Universe, or by SCUBA at $z\sim2-3$.  As such, \boofn\ provides a unique opportunity to study in detail the properties of a dusty starburst galaxy at moderate redshift.  This object is also interesting since the Near-IR and IRS spectroscopy verify that it lies in the relatively unexplored spectroscopic desert.  For future studies, observations of the CO molecular lines would be valuable in determining the gas content in which the stars are forming.  More urgently, high spatial resolution optical and Near-IR imaging is essential to understand and correct for the lensing characteristics of this galaxy.

\acknowledgments
We thank Shri Kulkarni and Derek Fox for generously observing this object with LRIS-B which inspired the more extensive optical spectroscopy effort with DEIMOS.  Advice from Scott Chapman and Ian Smail greatly improved this manuscript.  The work of MB, PE, and DS was carried out at the Jet Propulsion Laboratory, under contract with NASA.   AWB is supported by NSF AST-0208527, the Sloan Foundation and the Research Corporation.  This work was supported in part by the National Science Foundation through its support of the National Optical Astronomy Observatory, which is operated by the Association of Universities for Research in Astronomy, Inc. (AURA).  We thank the NDWFS team for the NDWFS data products used in this work, and NOAO for supporting the NDWFS.  The authors also wish to acknowledge the very significant cultural role that the summit of Mauna Kea has for the indigenous population of Hawaii.

Facilities: {Spitzer(MIPS)}, {Spitzer(IRAC)}, {CSO(SHARC-II)}, {Keck(NIRSPEC)}, {Keck(DEIMOS)}, {JCMT(SCUBA)}, {KPNO(MOSAIC-1)}

\end{document}